\documentclass[prl,twocolumn,superscriptaddress]{revtex4}
\usepackage{amssymb,amsmath,amsthm,graphicx}
\usepackage{latexsym}
\usepackage{epstopdf}
\usepackage{bm}
\usepackage[]{hyperref} 

\def\be{\begin{equation}}
\def\ee{\end{equation}}
\def\beq{\begin{eqnarray}}
\def\eeq{\end{eqnarray}}

\pdfoutput=1

\begin{document}

\def\lsim{\mathrel{\rlap{\lower4pt\hbox{\hskip1pt$\sim$}}
    \raise1pt\hbox{$<$}}}
\def\gsim{\mathrel{\rlap{\lower4pt\hbox{\hskip1pt$\sim$}}
    \raise1pt\hbox{$>$}}}
\def\be{\begin{equation}}
\def\ee{\end{equation}}
\def\bea{\begin{eqnarray}}
\def\eea{\end{eqnarray}}
\newcommand{\der}[2]{\frac{\partial{#1}}{\partial{#2}}}
\newcommand{\dder}[2]{\partial{}^2 #1 \over {\partial{#2}}^2}
\newcommand{\dderf}[3]{\partial{}^2 #1 \over {\partial{#2} \partial{#3}}}
\newcommand{\eq}[1]{Eq.~(\ref{eq:#1})}
\newcommand{\dd}{\mathrm{d}}

\title{Hidden singularities and closed timelike curves in a proposed dual for Lifshitz-Chern-Simons gauge theories}

\author{Keith Copsey}
\email{kcopsey@perimeterinstitute.ca}
\affiliation{ Perimeter Institute for Theoretical Physics, Waterloo, Ontario N2L 2Y5, Canada}
\author{Robert B. Mann}
\email{rbmann@sciborg.uwaterloo.ca}
\affiliation{ Perimeter Institute for Theoretical Physics, Waterloo, Ontario N2L 2Y5, Canada}
\affiliation{Department of Physics \& Astronomy, University of Waterloo, Waterloo, Ontario N2L 3G1, Canada}

\begin{abstract}
We point out that the metrics recently proposed by K. Balasubramanian and J. McGreevy \cite{BalaMcGreevyLifshitz} as gravitational duals to Lifshitz Chern-Simons gauge theories contain both a hidden null singularity and a region of closed timelike curves accessible to asymptotic observers.   Like the singularity in the original Liftshitz spacetime given by Kachru, Liu, and Mulligan, this  singularity does not include large $\alpha'$ or $g_s$ corrections and hence appears to be singular in string theory as well as classically.
\end{abstract}

\maketitle


There has recently been a significant amount of interest in finding gravitational duals to field theories with anisotropic scaling between time and space directions known as Lifshitz theories.  The most obvious possibility for such duals are spacetimes with corresponding anisotropies.   The simplest such solutions were given by Kachru, Liu, and Mulligan \cite{KLM} but those solutions suffer from a null curvature singularity that can not be cured by $\alpha'$ or $g_s$ effects \cite{CopseyMannLifshitz}.   Further, it has recently been argued \cite{HorowitzBenson} that strings propagating in such a background become infinitely excited and so this solution should be regarded as singular in string theory as well as classically.  It seems remarkably difficult for the spacetime to avoid such a singularity--the obvious candidates to smooth out the singularity in \cite{KLM} either fail to exist or also posses a null curvature singularity \cite{CopseyMannLifshitz}.

Very recently a remarkably simple solution of type IIB supergravity has been proposed by Balasubramanian and McGreevy \cite{BalaMcGreevyLifshitz} as a dual to a Lifshitz Chern-Simons theory:
\begin{eqnarray} \label{McGreevymet1}
ds^2 &=& L^2 \Big( \frac{2 dx_3 dt + d\vec{x}^2 + dr^2}{r^2} + f(r) dx_3^2 + d\Omega_5^2\Big) \nonumber \\
f(r) &=& f_0 \Big(1 - \frac{r^2}{{r^2_\star}} \Big) \nonumber \\
F_5 &=& 2 L^4 (\Omega_5 + \star \Omega_5) \,,  \,   \, \,  C_0 = \frac{Q x_3}{L_3} \,,  \,   \, \, \Phi = \Phi_0
\end{eqnarray}
where $d\Omega_5^2$ is the unit metric on $S_5$, $C_0$ the RR axion, and
\be
f_0 = \frac{Q^2 e^{2 \Phi_0}}{4 L_3^2}
\ee
In the above, $x_3$ is taken to be a compact direction with period $L_3$.  The above metric approaches $AdS_5 \times S_5$, except with a non-normalizable deformation (from $f(r)$) and so has an asymptotic anisotropic scaling symmetry: $t \rightarrow \lambda^2 t, \, \,  r \rightarrow \lambda r, \, \, \vec{x} \rightarrow \lambda\vec{x}, \, \, x_3 \rightarrow x_3 $.  

The assertion in \cite{BalaMcGreevyLifshitz} is that the spacetime smoothly ends at $r = r_\star$ where $f(r)$ vanishes.  However as $f(r)$ becomes small the spacetime is increasingly well approximated by $AdS_5 \times S_5$ in Poincare slicing, which certainly does not end at any nonzero $r$.   More concretely, defining
\be
x_3 = \frac{z + \tau}{\sqrt{2}}, \, \,\, \, \, \,  \, t = \frac{z - \tau}{\sqrt{2}}
\ee
we see that as $r \rightarrow r_{\star}$
\be
ds^2 \rightarrow  L^2 \Big( \frac{-d\tau^2 + dz^2 + d\vec{x}^2 + dr^2}{r^2} + d\Omega_5^2\Big) 
\ee
and it should be clear one may travel harmlessly through the surface $r = r_\star$ to larger values of $r$ (i.e. deeper into the interior of the spacetime in these coordinates).  We will, however, confirm this argument with an analysis of the motion of geodesics below.

Given the number of symmetries in the above metric it is easy to solve for the motion of geodesics in this background.  
Since the spacetime admits the Killing  vectors $\partial/{\partial t}$, $\partial/{\partial x_3}$, and $\partial/{\partial x_i}$(for $i = 1, 2$) any geodesic has a conserved energy
\be
E = - \frac{dx^\mu}{d \lambda} \Big(\frac{\partial}{\partial t} \Big)_{\mu}
\ee
and conserved momenta
\be
p_3 =  \frac{dx^\mu}{d \lambda} \Big(\frac{\partial}{\partial x_3} \Big)_{\mu}
\qquad
p_i =  \frac{dx^\mu}{d \lambda} \Big(\frac{\partial}{\partial x_i} \Big)_{\mu}
\ee
and so given a null ($k = 0$) or timelike ($k= 1$) geodesic one has an effective potential
\be
0 = \dot{r}^2 + V_{eff}
\ee
where
\be
V_{eff} = k \frac{r^2}{l^2} -E^2 \frac{r^6}{l^4} f(r) -(2 E p_3 - p_1^2 - p_2^2)  \frac{r^4}{l^4}
\ee
which we illustrate in Figure 1.
\begin{figure}
    
\includegraphics[scale= 1.10]{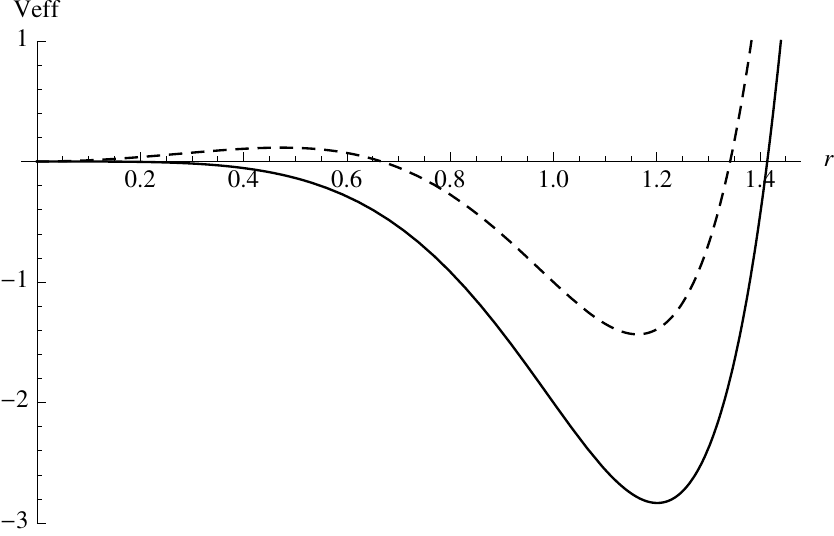}
	\caption{Effective potential for null (solid) and timelike (dashed) geodesics with $E = p_3 = f_0 = l = r_{\star} = 1$ and $p_1 = p_2 = 0$}
	\label{Vpotfig}
	\end{figure}
Just as in asymptotically AdS spaces,  null geodesics with a relatively large amount of momentum along the transverse ($\vec{x}$) directions and all timelike geodesics turn around at some finite $r$ and never reach infinity ($r = 0$) while null geodesics with a relatively small transverse momentum reach infinity in finite coordinate time $t$ (although, as usual, in infinite affine parameter $\lambda$).  In the interior of the spacetime, provided that $E$ and $p_3$ are nonzero and their product is large enough so that
\be
Q_0 \equiv 2 E p_3 - p_1^2 - p_2^2 - k \frac{l^2}{r_{\star}^2} > 0
\ee
then geodesics traveling towards the interior of the spacetime proceed smoothly through $r = r_{\star}$ to larger $r$.  In particular, near this point
\be
r - r_{\star} \approx \pm \frac{r_{\star}^2}{l^2} \sqrt{Q_0} (\lambda - \lambda_0)
\ee
where $r(\lambda_0) = r_{\star}$ and the sign is chosen depending on whether one is considering a radial ingoing or a radial outgoing geodesic.   

For sufficiently large $r$, $V_{eff}$ always becomes positive and so no geodesics get to infinite $r$.  The point $r_0$ where $V_{eff}$ goes through a zero, and hence the maximum possible $r$ for a given geodesic, is given for null geodesics with $Q_0 > 0$ by
\be
r_0^2 = \frac{r_{\star}^2}{2}\Big(1+\sqrt{1+\frac{4 Q_0}{f_0} \frac{l^2}{r_{\star}^2}} \Big)
\ee
and hence $r_0 > r_{\star}$ as long as $Q_0 > 0$.   Again assuming $Q_0 > 0$, for null geodesics the extremum of $V_{eff}$ occurs at
\be
r_1^2 = \frac{3 r_{\star}^2}{8}\Big(1+\sqrt{1+\frac{32 Q_0}{9 f_0} \frac{l^2}{r_{\star}^2}} \Big)
\ee
and $r_1$ may be larger or smaller than $r_{\star}$ depending on the size of $Q_0$ one chooses.   For timelike geodesics the analogous statements to the above require solving a cubic equation. For our purposes it will be sufficient to note that if one takes $E$ fixed and an increasingly large $p_3$ for timelike geodesics
\be \label{tlim1}
r_0^2 = \frac{r_{\star}}{l^2}  \frac{\sqrt{Q_0}}{E \sqrt{f_0}} \Big(1+ \mathcal{O}(Q_0^{-1/2}) \Big)
\ee
and
\be
r_1^2 = \frac{r_{\star}}{\sqrt{2} l^2}  \frac{\sqrt{Q_0}}{E \sqrt{f_0}} \Big(1+ \mathcal{O}(Q_0^{-1/2}) \Big)
\ee

Presuming one periodically identifies $x_3$, as in \cite{BalaMcGreevyLifshitz}, then since $f(r) < 0$ for $r > r_{\star}$
\be
\Big(\frac{\partial}{\partial x_3} \Big)^a \Big(\frac{\partial}{\partial x_3} \Big)_a = g_{x_3 x_3} = f(r) < 0
\ee
and for $r > r_{\star}$ one enters a region of closed timelike curves (CTCs).   Further, as discussed above, geodesics can travel from the asymptotic region into the CTC region or, if one prefers, spend an arbitrarily long period of time in the CTC region by choosing a geodesic that sits at or near a minimum of the effective potential.   Hence there is no sense in which these closed timelike curves are hidden or inaccessible.   At least at the level of supergravity, these could be avoided by not periodically identifying $x_3$, although it is less clear that such a spacetime should have any role as the desired dual \cite{BalaMcGreevyLifshitz}.

Even if one were content to live with closed timelike curves, this spacetime also contains a singularity as $r \rightarrow \infty$.  The simplest way to see this singularity is to examine the components of the Riemann tensor in a parallelly propagated orthonormal frame (PPON).   Taking one basis vector $e_0$ parallel to a timelike geodesic that, for the sake of simplicity, we will take to have no transverse momentum ($p_1 = p_2 = 0$) and a second basis vector $e_1$ proportional to $\partial/{\partial x_1}$ one finds
\begin{eqnarray}
R_{0 1 0 1} &\equiv& R_{\alpha \beta \gamma \delta} (e_0)^{\alpha} (e_1)^{\beta} (e_0)^{\gamma} (e_1)^{\delta} \nonumber \\
 &=& \frac{1}{l^2} + E^2 f_0 \frac{r^4}{l^4} \Big(1 - 2\frac{r^2}{r^2_\star} \Big) 
 \end{eqnarray}
Since, as noted above (\ref{tlim1}), there are geodesics that reach arbitrarily large $r$ for sufficiently large $p_3$ there are geodesics that extend into regions of arbitrarily high curvature.  Note that the norm of all vectors at a constant $r$ and at a fixed point in the $S_5$ goes to zero as $r \rightarrow \infty$ and so, just as with the Poincare horizon, the surface $r \rightarrow \infty$ is a null surface.   Hence, here one has a null curvature singularity.

One of the remarkable properties of null curvature singularities is that they do not necessarily make any curvature invariant diverge.  The null curvature singularity in the original Lifshitz spacetime \cite{KLM}, as well as in singular gravitational plane waves \cite{HorowitzSteif} and a variety of other examples \cite{HorowitzRossNaked}, is of this type.  This is a particularly useful property from the point of view of string theory since as long as all curvature invariants remain small $\alpha'$ corrections remain negligable and if, in addition, the dilaton never becomes large, as in the solution under consideration here (\ref{McGreevymet1}), the supergravity approximation remains a good one and the solution apparently should be regarded as singular in string theory as well as classically.   

Aside from the case \cite{HorowitzSteif} where one can make a simple symmetry argument showing that all curvature invariants vanish, perhaps the simplest way to establish that all curvature invariants never diverge is to show there is a basis where all the components of the Riemann tensor are everywhere finite.  Then let us consider a set of orthonormal basis vectors
$$
\tilde{e}_0 = \alpha(r) dt + \beta(r) dx_3, \, \, \, \, \tilde{e}_3 = \gamma(r) dt_a + \delta(r) dx_3 
$$
\be
\tilde{e}_i = \frac{l}{r} dx_i , \, \, \, \, \tilde{e}_4 = \frac{l}{r} dr , \, \, \, \, \tilde{e}_a = \Omega_a
\ee
where $\Omega_a$ are a set of  basis vectors for the unit $S_5$.  Demanding that $\tilde{e}_0$ is a unit timelike vector and $\tilde{e}_1$ a unit spacelike vector and $\tilde{e}_0$ and $\tilde{e}_1$ are orthogonal fixes three of the four functions $(\alpha, \beta, \gamma, \delta)$ (up to an overall sign that is fixed if, as we do, $\alpha$ and $\gamma$ are taken to have the same sign). Leaving $\alpha(r)$ free for the moment we find  the non-trivial components of the Riemann tensor contracted into these basis vectors to be
\begin{eqnarray}
\tilde{R}_{0 3 0 3} &=& \frac{1}{l^2} \, \, \, \, \,  \, \, \, \, \, \, \tilde{R}_{0 i 3 i} = \alpha^2 f_0 \frac{r^4}{l^4}  \Big(1 - 2 \frac{r^2}{r^2_{\star}} \Big)  \nonumber \\
\tilde{R}_{0 i 0 i} &=& \frac{1}{l^2} + \alpha^2 f_0 \frac{r^4}{l^4} \Big(1 - 2 \frac{r^2}{r^2_{\star}} \Big) \nonumber \\
\tilde{R}_{3 i 3 i} &=& -\frac{1}{l^2} + \alpha^2 f_0 \frac{r^4}{l^4} \Big(1 - 2 \frac{r^2}{r^2_{\star}} \Big) \nonumber \\
\tilde{R}_{0 4 0 4} &=& \frac{1}{l^2} + \frac{4 f_0 \alpha^2 r^6}{l^4 r_c^2} \,  \, \, \, \, \, \, \, \, \, \tilde{R}_{0 4 3 4} = \frac{4 f_0 \alpha^2 r^6}{l^4 r_c^2} \nonumber \\
\tilde{R}_{3 4 3 4} &=& -\frac{1}{l^2} + \frac{4 f_0 \alpha^2 r^6}{l^4 r_c^2} \nonumber \\
\tilde{R}_{i j i j} &=&  \tilde{R}_{i 4 i 4} = -\frac{1}{l^2}
\end{eqnarray}
and choosing a suitable $\alpha(r)$, e.g.
\be
\alpha(r) = (1+\frac{r^2}{l^2})^{-3/2}
\ee
all of the components of the Riemann tensor in the $\tilde{e}$ basis are finite everywhere and hence no curvature invariant, and hence no $\alpha'$ correction, ever becomes large.

Hence we conclude that this null curvature singularity can not resolved by either $\alpha'$ or $g_s$ effects.   It is not immediately clear whether test strings propagating on this background would become infinitely excited; unfortunately there does not seem to be any obvious way to transform the near singularity region into plane wave coordinates and immediately apply the results of \cite{HorowitzBenson}.   From the point of view of diverging tidal forces infinite string excitation would not be a surprising result, but on the other hand this singularity has the rather unusual feature of repelling both timelike and null geodesics, so it is also plausible that strings might be simply repelled away from the singularity without exciting many high frequency modes.   In any case, any stringy, or more general Planck scale, physics will be visible to distant observers as noted before.  The stability of these solutions remains an open question, although it seems likely if one enforces the periodic identification of $x_3$ one should expect instabilities in at least the CTC region $r > r_{\star}$.

\vskip 1cm

\centerline{\bf Acknowledgements}
\vskip .2cm
 This work was supported in part by the Natural Sciences \& Engineering Research
Council of Canada.   Research at Perimeter Institute is supported by the Government of Canada through Industry Canada and by the Province of Ontario through the Ministry of Research \& Innovation.


\end{document}